\documentclass[runningheads]{llncs}
\usepackage{marvosym}
\usepackage[T1]{fontenc}
\usepackage{graphicx}
\usepackage{fontawesome}
\usepackage{subfigure}
\usepackage{hyperref}
\usepackage{xurl}
\usepackage[flushmargin,hang]{footmisc}
\begin{document}
%
\title{When BBR Meets Live Streaming}

%
%

\newcommand{\cormail}{\textsuperscript{%
    (\fontsize{8.5pt}{9pt}\selectfont \kern0.15em {\Letter}
}}

\author{
    Xu Yan\inst{1}\and
    Tong Li\inst{1,4}\textsuperscript{\Letter} \and
    Bo Wu\inst{2}\and
    Cheng Luo\inst{2}\and
    Jiuxiang Zhu\inst{1}\and
    Laizhong Cui\inst{3}
}
\authorrunning{Yan et al.}
%
\institute{Renmin University of China, Beijing, China \\
\email{\{yanxu2,tong.li,2024103876\}@ruc.edu.cn}\\ \and
Tencent Technologies, Beijing, China\\
\email{\{brynwu,lancelotluo\}@tencent.com}\\ \and
Shenzhen University, Guangdong, China\\
\email{cuilz@szu.edu.cn} \\ \and
State Key Laboratory of Internet
Architecture, Tsinghua University, Beijing, China
}


%
\maketitle              

\let\thefootnote\relax\footnotetext{
This work is partially supported by the National Natural Science Foundation of China (62572473, 62441230, U23B2026, and 62372305), the Scientific Research Innovation Capability Support Project for Young Faculty (SRICSPYF-ZY2025001),  the funding from State Key Laboratory of Internet Architecture (HLW2025ZD17), the Guangdong Basic and Applied Basic Research Foundation (2024B1515040012), and  the  funding from Tencent Basic Platform Technology Rhino-Bird Focused Research Program.}

\begin{abstract}

Recently, industrial pioneers like Amazon, Tencent, ByteDance, and Huawei have been adopting BBR as their congestion control algorithm for live-streaming applications, including TikTok Live. However, BBR, originally crafted for bulk data transmission, faces multiple challenges in live-streaming scenarios.
In this paper, we first explore two key issues associated with BBR due to inaccurate bandwidth estimation in live-streaming scenarios: (i) BBR cannot easily exit its startup phase, resulting in a fierce self-inflicted loss. (ii) BBR sends data at a lower rate than the available bandwidth during its stable phase. We then propose BBR-Copilot, an auxiliary congestion control component that cooperates with BBR, making BBR better adapt to live-streaming scenarios. BBR-Copilot allows for proactively generating accurate bandwidth measurement samples by smartly creating and sending extra data. We implement the BBR-Copilot prototype upon QUIC and evaluate it via testbed. Experimental evaluation results show that BBR-Copilot effectively enhances BBR's performance in live-streaming scenarios.

\end{abstract}

\newcommand{\name}{BBR-Copilot}

\section{Introduction}
\label{introduction_sec}
\vspace{-2mm}
With the rapid development of the Internet, live-streaming services such as TikTok Live~\cite{tiktok2024}, YouTube Live~\cite{youtubelive2024}, and Twitch~\cite{twitch2024} have become essential elements of our daily entertainment. Users tend to prefer live-streaming platforms that offer a better viewing experience for watching live streams for longer periods. For live-streaming service providers and Content Delivery Network (CDN) service providers, having more users and longer viewing times often translates to higher profits. Therefore, improving user's viewing experience has become a significant challenge for both live-streaming service providers and CDN service providers.



As a state-of-the-art congestion control algorithm, BBR (Bottleneck Bandwidth and Round-trip propagation time) \cite{CardWell2016bbr} has been demonstrated to perform excellently and has been widely adopted (in \S \ref{bbr_congestion_control_sec}). During its \emph{Startup phase}, BBR rapidly increases its data-sending rate to quickly fill available bandwidth. In the \emph{ProbeBW phase} (a.k.a, stable phase), BBR continuously monitors bottleneck bandwidth and Round-Trip Time (RTT) to reduce queuing delay in the intermediate buffer while maintaining high link utilization. Due to BBR's potential benefits in terms of Quality of Service (QoS) for data transmission and the user viewing experience, industrial pioneers such as Amazon~\cite{amazon2024}, Tencent\cite{tencent2024}, ByteDance~\cite{bytedance2024} and Huawei~\cite{huaiwei2024} have chosen BBR as their congestion control algorithm in cloud services or applications.


However, BBR encounters several issues due to lacking accurate bandwidth measurement samples in live-streaming scenarios. In live-streaming scenarios, the upper-layer application generates a frame of video periodically to be delivered to the transport layer. If the streaming bitrate is lower than the actual available bandwidth, it frequently causes the transport layer to enter a phase where it can send data but has no data to send, which is called the application-limited phase~\cite{cheng2022estimation} (in \S \ref{transmission_characteristics_sec}). This frequent entry into the application-limited phase results in many inaccurate bandwidth measurement samples (in \S \ref{bandwidth_sample_sec}), which in turn causes BBR to encounter two key issues. \textbf{Firstly}, BBR fails to exit the Startup phase. During the Startup phase, the sender tends to send excessive data, leading to fierce losses. \textbf{Secondly}, even when BBR exits the Startup phase and turns into the ProbeBW phase, BBR cannot accurately estimate the available bandwidth, leading to long frame completion time. A high RTT prominently highlights these issues (in \S \ref{problems_with_bbr_sec}).


To tackle the issues BBR encounters in live-streaming scenarios, it is natural to consider transforming live streaming into bulk data transmission by sending extra data. This approach prevents the transport layer from entering the application-limited phase and ensures that bandwidth measurement samples are accurate. However, sending extra data at all times would incur excessive transmission costs, which is unacceptable. Therefore, we must exercise meticulous control over the sending of extra data (in \S \ref{insight_and_challenge_sec}).

In this paper, we propose \name{} (in \S \ref{design_overview_sec}), which can enhance the performance of BBR in live-streaming scenarios with transmission costs as low as possible. Specifically, when the \emph{padding controller} (in \S \ref{extra_control_module_sec}) of \name{} detects that BBR is currently in the bandwidth probing phase (i.e., pacing\_gain is greater than 1) and the current transport layer is in the application-limited phase, it will control the \emph{data generator} (in \S \ref{extra_data_generation_sec}) to generate and send extra data, enabling the bandwidth measurement mechanism to generate accurate bandwidth measurement samples when BBR requires them.
Therefore, in the Startup phase, BBR can choose to exit the Startup phase based on the latest bandwidth measurement samples and the transition conditions of its state machine, avoiding the potentially large number of packet losses in a shallow buffer scenario. In the ProbeBW phase, \name{} can increase the accuracy of bandwidth measurement samples, helping BBR to adjust the sending rate more accurately based on the latest bandwidth measurement samples.

We implement \name{} prototype upon QUIC protocol~\cite{lsquic2024,langley2017quic,iyengar2021quic} with just 200 lines of code
and conduct experiments on our testbed (in \S \ref{testbed_structure}). The evaluation results are as follows. (i) \name{} can increase the ratio of BBR exits from Startup to 100\%, effectively reducing the retransmission ratio by 43.3\% in live-streaming scenarios (in \S \ref{evaluation_startup_sec}).
(ii) \name{} helps BBR to reduce the Root Mean Square Error (RMSE) of the true bandwidth and the bandwidth measurement samples by 86.1\% and adjust the sending rate to match the available bandwidth in live-streaming scenarios (in \S \ref{evaluation_probebw_sec}).

\vspace{-3mm}

\section{Background and Motivation}
\label{background_sec}


\vspace{-3mm}

In this section, we first introduce the BBR congestion control algorithm (in \S \ref{bbr_congestion_control_sec}). Then, we highlight the recurrent occurrence of application-limited phases in live-streaming scenarios, contrasting sharply with bulk data transmission (in \S\ref{transmission_characteristics_sec}). Furthermore, we dissect BBR's bandwidth estimation process and diagnose estimation inaccuracy stemming from application-limited phase (in \S \ref{bandwidth_sample_sec}). Finally, we present two challenges faced by BBR in live-streaming scenarios (in \S \ref{problems_with_bbr_sec}).


\begin{figure}[!t]
    \includegraphics[width=11cm]{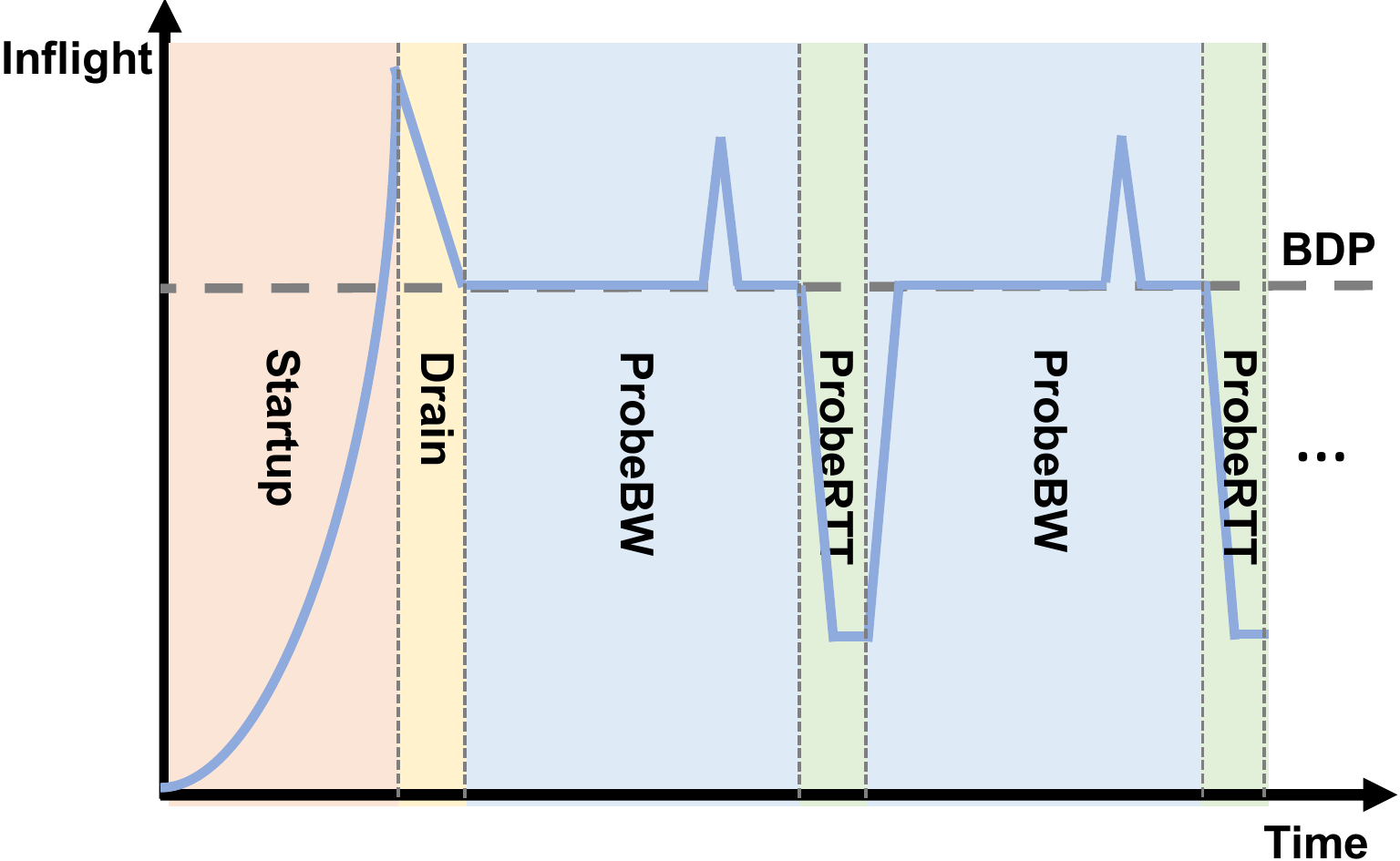}
    \vspace{-0.2cm}
    \caption{BBR congestion control curve}
    \label{bbr}
    \vspace{-0.6cm} 
\end{figure}

\vspace{-3mm}
\subsection{BBR Overview}
\label{bbr_congestion_control_sec}
BBR is a new-type congestion control algorithm proposed by Google in 2016\cite{CardWell2016bbr}. BBR was originally designed for file transfer to optimize transfer performance. Traditional congestion control algorithms are mostly based on the signal of packet loss or RTT. However, BBR is based on active detection. 
The kernel method of BBR is an estimation of RTprop (round-trip propagation
time) and BtlBw (bottleneck bandwidth) to maximize transport throughput and minimize delivery delay. Bandwidth-Delay Product (BDP) is the product of the BtlBw and RTprop. It quantifies the amount of data that can be in transit in the network at any given time. BBR uses this value to adjust its sending rate and congestion window to maximize throughput while avoiding potential congestion.

As Figure \ref{bbr} shows, BBR operates in four main phases: Startup, Drain, ProbeBW, and ProbeRTT. Startup is the initial phase, where BBR enables data to be sent at a rate that ramps up until available bandwidth is fully utilized. The Drain phase is carried out after the Startup phase, the sender stops sending follow-up packets until the in-flight data size (inflight) meets some condition (e.g., inflight $\leq$ BDP). 
When BBR exits the Drain phase, it enters the ProbeBW phase. BBR will periodically adjust the actual sending rate to probe BtlBw in this phase.  
Besides, BBR periodically enters the ProbeRTT phase, during which BBR maintains the inflight at a low level to precisely measure the RTprop.


BBR has revolutionized congestion control. Google, the developer of the BBR, has significantly enhanced the service performance of YouTube, Google.com, and Google Cloud Platform by using BBR. Cardwell et al. documented impressive results, such as a 2-25x decrease in video rebuffering rates on YouTube and up to a 14\% increase in throughput for Google.com search queries \cite{CardWell2016bbr}. BBR along with its Google variants are deployed at 22\% of classified websites, becoming one of the popular congestion control algorithms \cite{mishra2019great}. 


\vspace{-4mm}
\subsection{Application-limited Phase is Ubiquitous in Live Streaming}
\label{transmission_characteristics_sec}
\vspace{-1mm}
\begin{figure}[tbp]
    \centering
    \subfigure[Bulk data transmission]{
    \includegraphics[width=5.7cm, angle=0]{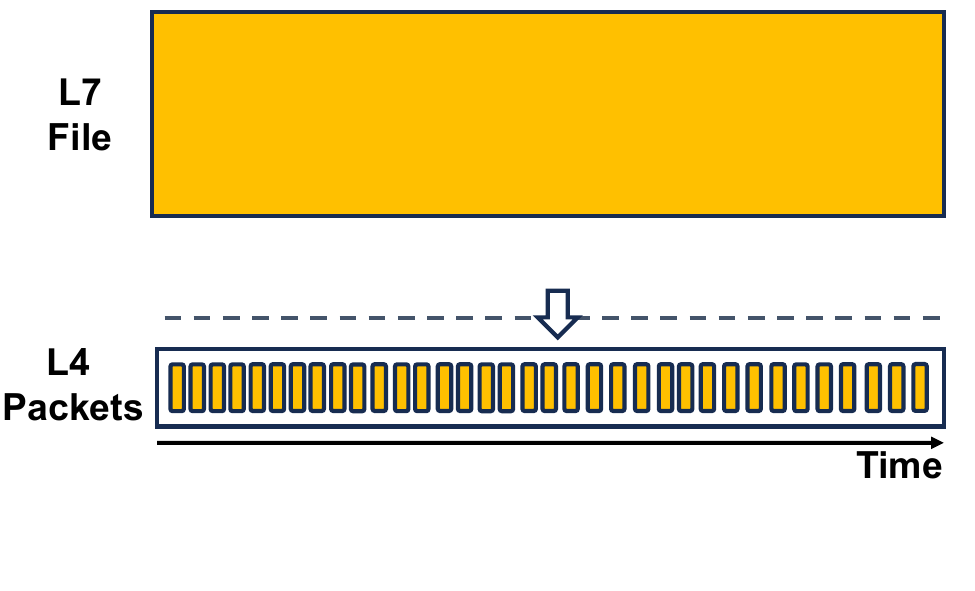}
    \label{send_character_bulk_data_fig}
    }
    \subfigure[Live steaming]{
    \includegraphics[width=5.7cm, angle=0]{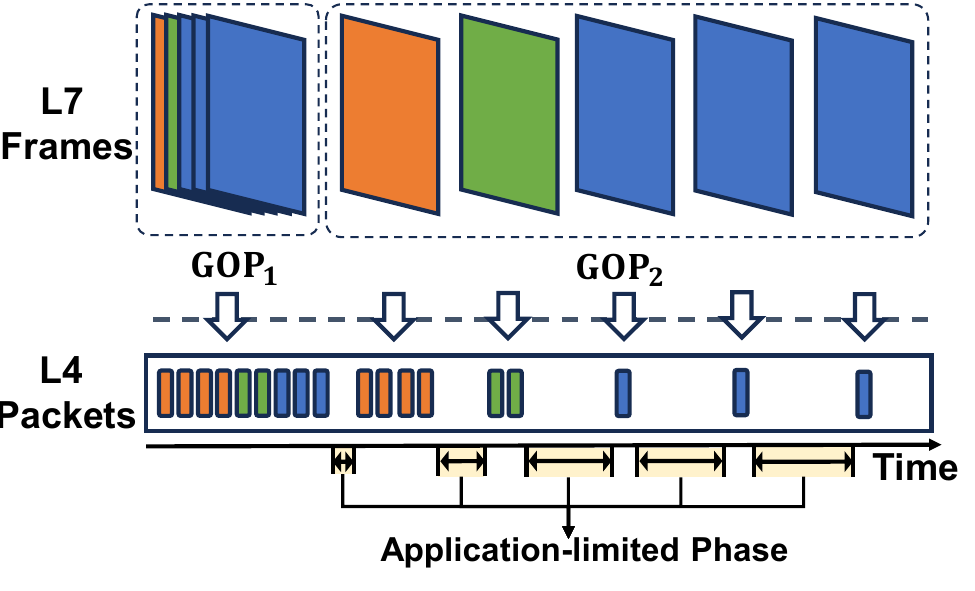}
    \label{send_character_live_streaming_fig}
    }
        \vspace{-0.2cm}
    \caption{Transmission characteristics.}
    \label{transmission_characters_fig}
    \vspace{-0.6cm}
\end{figure}

In traditional bulk data transmission, once the client's request is processed, the server's application layer submits the entire requested file to the transport layer, as shown in Figure \ref{send_character_bulk_data_fig}. Until the entire file is fully transmitted, there will be no instances where there is no data available to send in the send buffer. Therefore, when transmitting bulk data, the transport layer will hardly ever enter a phase where it can send data but has no data to send, which is called the application-limited phase.
However, the situation is significantly different in live-streaming scenarios.
As shown in Figure \ref{send_character_live_streaming_fig}, after the client's request is processed, to allow the client to buffer some frames and reduce the impact of network fluctuations, the server will first send the latest Group of Pictures (GOP) to the client, and then successively send the real-time generated video and audio frames. Therefore, when the connection is just established, the server has a lot of data to send, and the transmission characteristics at this time are similar to those of transmitting bulk data. However, after the data of this GOP is transmitted, the upper-layer application will generate a new video or audio frame to be sent by the transport layer at intervals according to the frame rate. If the current frame is sent before the next frame is generated, and the current congestion control algorithm allows us to continue sending data, the transport layer will fall into a situation where it can send data but has no data to send (i.e., it enters the application-limited phase). Since the bitrate of the live stream must be less than or equal to the true bandwidth, otherwise the client will not be able to receive the corresponding frames in time and will frequently experience freezing, the transport layer of the sender will frequently enter the application-limited phase in live-streaming scenarios~\cite{yan2023too}.


\subsection{BBR Bandwidth Estimation Suffers From Application-limited Phase}
\label{bandwidth_sample_sec}

\begin{figure}[tbp]
    \centering
    \subfigure[Bulk data transmission]{
    \includegraphics[width=5.7cm, angle=0]{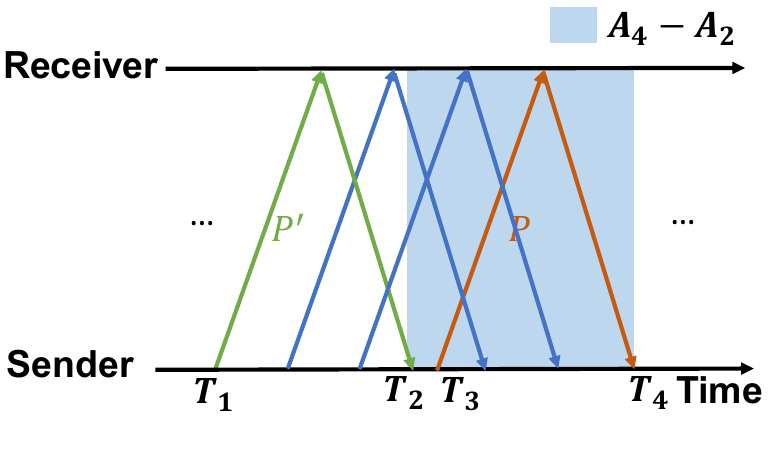}
    \label{bandwidth_measurement_bulk_data_fig}
    }
    \subfigure[Live streaming]{
    \includegraphics[width=5.7cm, angle=0]{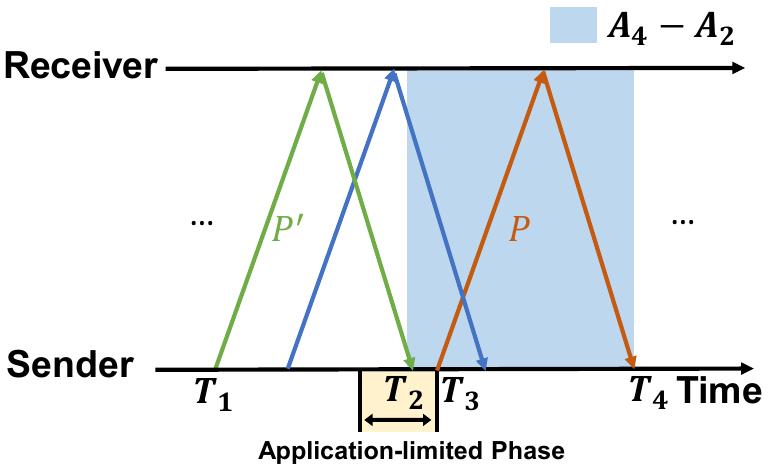}
    \vspace{-0.2cm}
    \label{bandwidth_measurement_applimited_data_fig}
    }
    \vspace{-0.2cm}
    \caption{Bandwidth measurement sample calculation.}  \label{bandwidth_measurement_sample_calculation_fig}
    \vspace{-0.4cm}
\end{figure}

\vspace{-3mm}
\vspace{1ex} \noindent \textbf{Bandwidth estimation in BBR is based on bandwidth measurement samples.} BBR estimates bandwidth by taking the maximum value of all bandwidth measurement samples computed from packets within a specific time window. The sample calculation procedure is as follows: As shown in Figure 
\ref{bandwidth_measurement_bulk_data_fig}, when packet $P$ is sent, we record the time $T_{3}$. When $P$ is acknowledged, we record the time $T_{4}$ when $P$ is acknowledged and the total amount of data $A_{4}$ acknowledged from the start of the connection to the time $T_{4}$. In addition, we also record some information about the \textit{most recently acknowledged packet $P^{'}$ when $P$ is sent}. This includes the time $T_{1}$ when $P^{'}$ is sent, the time $T_{2}$ when $P^{'}$ is acknowledged, and the total amount of data $A_{2}$ acknowledged from the start of the connection to the time $T_{2}$. When $P$ is acknowledged, we can calculate the average sending rate ($send\_rate$) and the average acknowledging rate ($ack\_rate$) of the packets sent from the time $T_{1}$ to $T_{3}$ according to Equation (\ref{send_rate_equ}) and Equation (\ref{ack_rate_equ}). 
\begin{equation}\label{send_rate_equ}
        {send\_rate} = \frac{{A}_{4} - {A}_{2}} {{T}_{3} - {T}_{1}}
\end{equation}
\begin{equation}\label{ack_rate_equ}
        ack\_rate =  \frac{A_{4} - {A}_{2}} {{T}_{4} - {T}_{2}}
\end{equation}

\noindent Ultimately, upon acknowledging $P$, the bandwidth measurement sample is derived from the lower of $send\_rate$ and $ack\_rate$.


\vspace{1ex} \noindent \textbf{The application-limited phase causes the bandwidth measurement samples to fail to reflect the underlying true bandwidth.} 
After the transport layer exits the application-limited phase, all bandwidth measurement samples generated by the packets sent from the time when the next packet is sent to the time when this packet is acknowledged are inaccurate. This is because the most recent acknowledged packet when these data packets were sent was sent before the transport layer entered the application-limited phase. According to Equations \ref{send_rate_equ} and \ref{ack_rate_equ}, since no data was sent when the transport layer was in the application-limited phase, this leads to the $send\_rate$ and $ack\_rate$ being less than the actual sending rate. Therefore, we cannot distinguish whether these bandwidth measurement samples are limited by the true bandwidth or the rate at which the upper-layer application submits data to the transport layer. So we mark these bandwidth measurement samples as inaccurate. In other words, when the transport layer enters the application-limited phase within the time of one RTT before a packet is sent, the bandwidth measurement sample generated by that packet becomes inaccurate. Therefore, the larger the RTT, the more inaccurate bandwidth measurement samples are caused by entering the application-limited phase. This is consistent with our experimental results in \S \ref{evaluation_sec}.

\begin{figure}[tbp]
    \centering
    \subfigure[High RTT]{
    \includegraphics[width=5.7cm, angle=0]{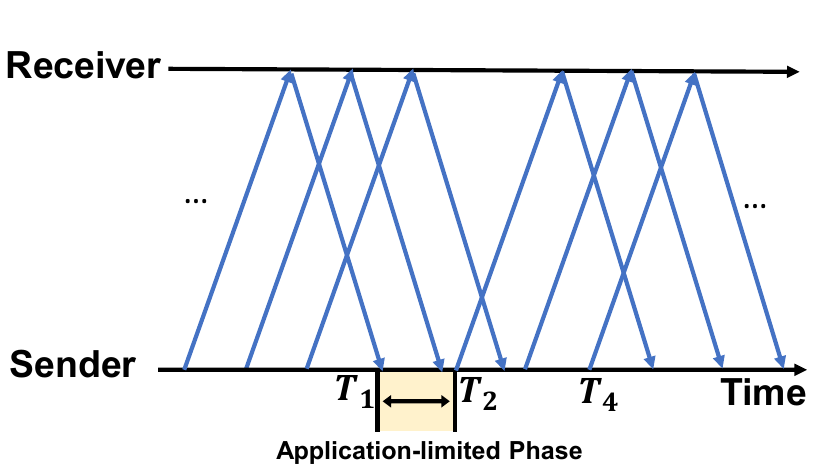}
    \label{bandwidth_measurement_live_streaming_long_rtt_fig}
    }
    \subfigure[Low RTT]{
    \includegraphics[width=5.7cm, angle=0]{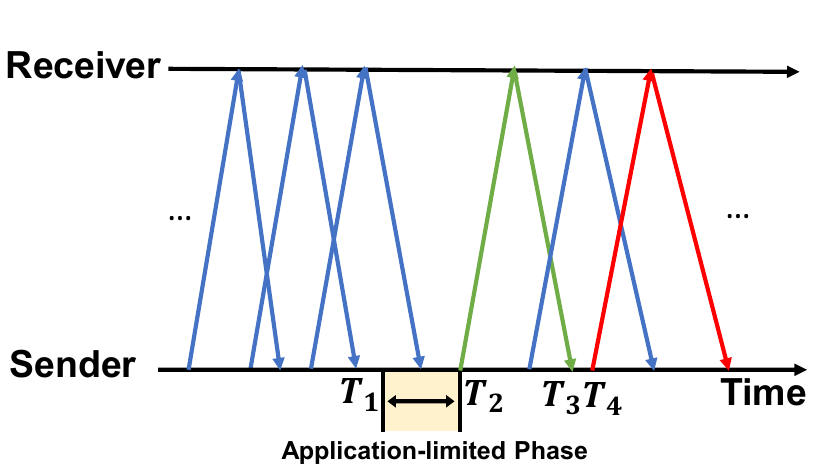}
    \vspace{-0.2cm}
    \label{bandwidth_measurement_live_streaming_short_rtt_fig}
    }
    \vspace{-0.2cm}
    \caption{Bandwidth measurement sample calculation in live streaming.}   \label{bandwidth_measurement_live_streaming_fig}
    \vspace{-0.4cm}
\end{figure}

%

Figure \ref{bandwidth_measurement_live_streaming_fig} provides an example illustrating how RTT affects the accuracy of bandwidth measurement samples in live-streaming scenarios. The RTT in Figure \ref{bandwidth_measurement_live_streaming_long_rtt_fig} is higher compared to Figure \ref{bandwidth_measurement_live_streaming_short_rtt_fig}, while all other conditions remain the same. At the time $T_1$, the transport layer enters the application-limited phase. In Figure \ref{bandwidth_measurement_live_streaming_long_rtt_fig}, this leads to all bandwidth samples measured between $T_2$ and $T_4$ being inaccurate. However, in Figure \ref{bandwidth_measurement_live_streaming_short_rtt_fig}, an accurate bandwidth sample can be calculated when the packet sent at time $T_4$ has been acknowledged.


\vspace{-3mm}

\subsection{Challenges When BBR Meets Live-Streaming}
\label{problems_with_bbr_sec}

\vspace{-2mm}
\vspace{1ex} \noindent \textbf{Challenge in Startup phase.}
In live-streaming scenarios, BBR does not easily exit the Startup phase. We conducted large-scale measurements on over 570,000 live-streaming flows in real-world networks and found that 88.4\% of them end with BBR remaining in the Startup phase. The specific reasons impeding BBR's smooth exit from Startup phase are examined in the following analysis. When the current round ends, BBR checks if the last bandwidth measurement sample is accurate. If it is inaccurate, BBR will not exit the Startup phase at this point. Otherwise, if the BtlBw estimated in the current round is 25\% higher than the BtlBw estimated in the previous round, a counter is reset to 0. Otherwise, the counter is incremented by 1. BBR exits the Startup phase when the counter reaches 3. In live-streaming scenarios, frequent transitions to the application-limited phase at the transport layer result in a large number of inaccurate bandwidth measurement samples. If the last bandwidth measurement sample within each round remains inaccurate, BBR can not exit the Startup phase. Based on the transmission characteristics mentioned in \S \ref{transmission_characteristics_sec}, in live-streaming scenarios, the server first sends a GOP to the client. If this data allows BBR to estimate a BtlBw close to the true bandwidth but not enough to exit the Startup phase, considering that BBR's sending rate is 2.885 times the BtlBw in the Startup phase, the sending rate will exceed the true bandwidth, leading to a high packet loss rate in shallow buffer scenarios. This phenomenon is exemplified in Figure \ref{the_optimization_effect_of_bbr_copilot_fig} (in \S\ref{evaluation_startup_sec}).

\vspace{1ex} \noindent \textbf{Challenge in ProbeBW phase.} 
During the ProbeBW phase, BBR periodically sets the pacing\_gain to 1.25, which means it will send data at a rate of 1.25 times the BtlBw to probe for more available bandwidth. When the true bandwidth increases, if a bandwidth measurement sample is generated during a period where the transport layer does not enter the application-limited phase and the pacing\_gain is set to 1.25, then the value of that bandwidth measurement sample will be greater than BBR's current estimated BtlBw, and BBR will update its BtlBw to the value of that bandwidth measurement sample, which allows the sending rate to continuously increase and eventually approach the true bandwidth. However, if all bandwidth measurement samples are taken during periods where the transport layer frequently enters the application-limited phase, resulting in an average sending rate within the measurement period that is lower than BBR's BtlBw, BBR will not use these bandwidth measurement samples to update its BtlBw. This leads to a problem where the sending rate does not increase even when the true bandwidth increases. Figure \ref{bandwidth_measurement_bbr_can_exit_startup_short_rtt_fig} (in \S\ref{evaluation_probebw_sec}) demonstrates a representative scenario of this phenomenon.

\vspace{-3mm}

\section{Design}
\label{design_sec}
\vspace{-3mm}

In this section, we first discuss the design rationale of how to allow for the generation of accurate bandwidth measurement samples (in \S \ref{insight_and_challenge_sec}). We then dive into the detailed design of \name{} by introducing the framework  (in \S \ref{design_overview_sec}) and its two key components  (in \S \ref{extra_control_module_sec} and \S \ref{extra_data_generation_sec}). 

\vspace{-2mm}

\subsection{Design Rationale}
\label{insight_and_challenge_sec}

To address the issues encountered by BBR mentioned above, one simple solution is to transform the live-streaming scenarios into bulk data transmission scenarios. This means generating extra data for transmission when the transport layer is about to enter the application-limited phase. By doing so, the transport layer will never enter the application-limited phase, ensuring that all bandwidth measurement samples are accurate. This naturally resolves the problems faced by BBR in live-streaming scenarios. However, this approach has a significant consequence: it results in the transmission of a large amount of extra data, wasting network resources and significantly increasing transmission costs.

Based on our understanding of BBR, we have identified that the two issues encountered by BBR in live-streaming scenarios are caused by a lack of accurate bandwidth measurement samples when the pacing\_gain is greater than 1. Therefore, we believe that only when the pacing\_gain of BBR is greater than 1 and the transport layer is about to enter the application-limited phase, extra data needs to be generated to avoid the transport layer being in an application-limited phase and ensure accurate bandwidth sampling, solving BBR's issue in live streaming.

\vspace{-2mm}
\subsection{Framework of BBR-Copilot}
\label{design_overview_sec}

\begin{figure}[tbp]
    \centering
    \includegraphics[width=9.0cm, angle=0]{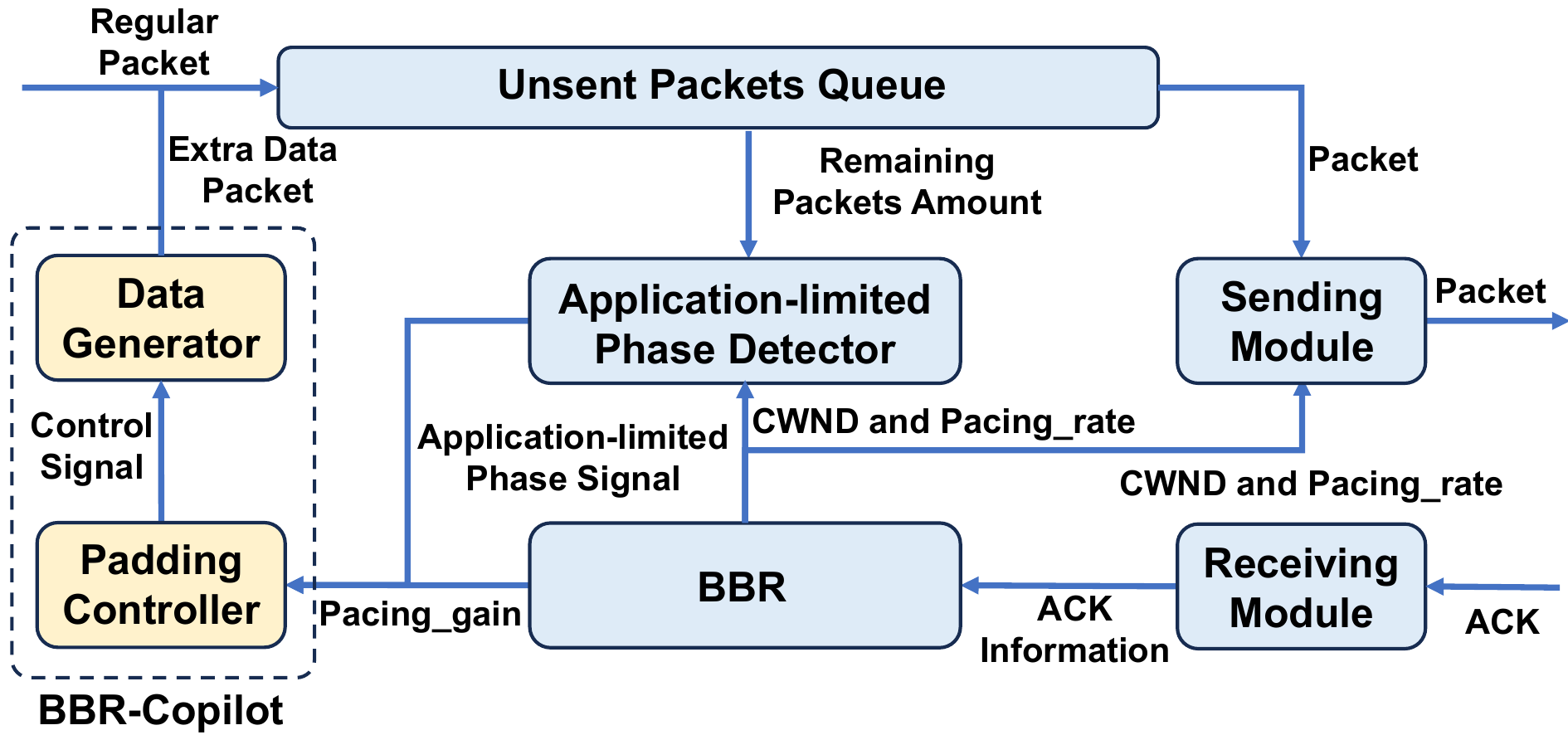}
        \vspace{-0.2cm}
    \caption{The framework of \name{} mechanism.}
    \label{bbr_copilot_overview_fig}
        \vspace{-0.6cm}
\end{figure}

As illustrated in Figure \ref{bbr_copilot_overview_fig}, \name{} primarily consists of two modules: The \emph{padding controller} and the \emph{data generator}. The padding controller is responsible for detecting whether the current conditions for generating extra data are met, thereby controlling the data generator. The data generator is responsible for receiving control signals from the padding controller and generating extra data to send when the conditions are met. When the sending module completes sending the packets that can be sent, \name{} checks if there is a need to generate extra data. It generates extra data when the transport layer is in an application-limited phase and BBR's pacing\_gain is greater than 1. Through this mechanism, we have achieved the function of the transport layer not entering an application-limited phase when BBR needs to probe bandwidth, enabling BBR to obtain accurate bandwidth measurement samples.



\vspace{-2mm}

\subsection{Padding Controller}
\label{extra_control_module_sec}
\vspace{-1mm}
After sending a packet, the padding controller starts checking if there is a need to generate extra data. Firstly, the padding controller queries the existing application-limited phase detector to determine if the transport layer is currently in an application-limited phase. If the application-limited phase detector determines that packets can be sent and detects that the unsent packet queue is empty, it returns an application-limited signal to the padding controller. Simultaneously, the padding controller obtains BBR's pacing\_gain. If the padding controller receives an application-limited signal and detects that the BBR's pacing\_gain is greater than 1, it sends the signal to the data generator to generate extra data. 


\vspace{-2mm}
\subsection{Data Generator}
\label{extra_data_generation_sec}
When the data generator receives the signal sent by the padding controller, it will generate a packet containing extra data, and then put this packet into the unsent packets queue, allowing it to be sent by the sending module. The extra data generated here is meaningless and is only used for assisting bandwidth estimation. After receiving this packet, the receiver will only return acknowledgment (ACK) and the data will not be submitted to the upper layer. This packet does not need to be resent after being lost, avoiding the problem of affecting subsequent normal data transmission.
\vspace{-3mm}


\section{Evaluation}
\label{evaluation_sec}

\vspace{-2mm}

In this section, we will introduce the experimental evaluation results of \name{}. We first introduce the specific architecture of the testbed (in \S \ref{testbed_structure}). Then, we evaluate the optimization effects of \name{} on the issue BBR encounters during the Startup phase (in \S \ref{evaluation_startup_sec}). Finally, we evaluate the optimization effects of \name{} on the issue BBR faces during the ProbeBW phase (in \S \ref{evaluation_probebw_sec}).

\vspace{-3mm}
\subsection{Architecture of the Testbed}
\label{testbed_structure}
We set up a test platform to evaluate \name{}. The test platform consists of a physical machine and two containers deployed on the physical machine. One of the containers simulates the source station, providing live streams. The physical machine simulates a CDN server, used to proxy the live streams, and \name{} is deployed on it. The other container simulates a client, which can send requests to the CDN server and obtain live-streaming data, and Mahimahi~\cite{ravi2015Mahimahi} is deployed on it to simulate the network environment between the CDN server and the client.

\vspace{-3mm}
\subsection{Optimization of \name{} in the Startup Phase}
\label{evaluation_startup_sec}

\begin{figure}[tbp]
    \centering
    \subfigure[Baseline]{
    \includegraphics[width=5.7cm, angle=0]{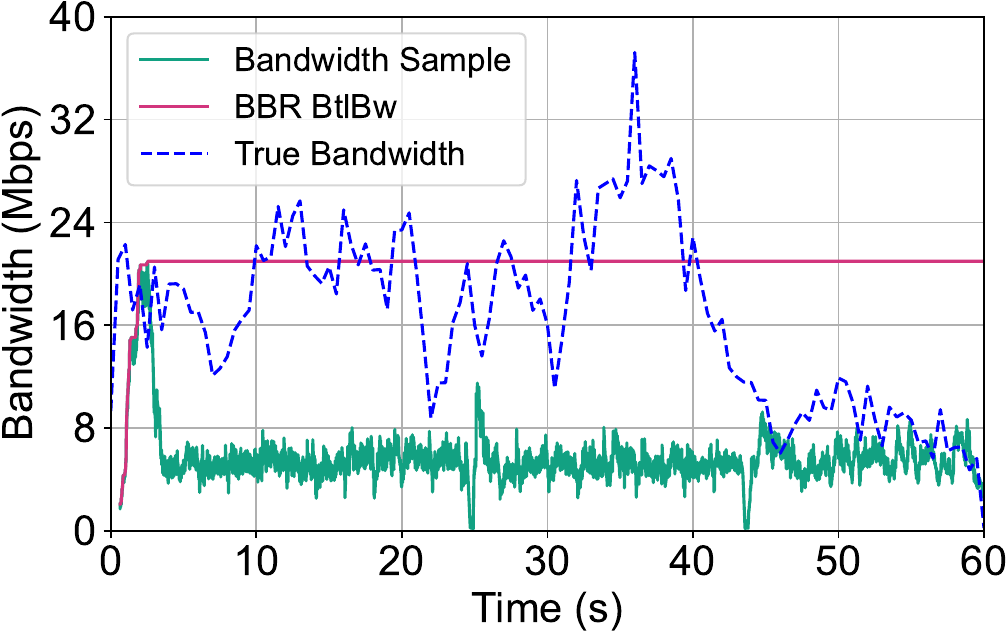}
    \label{bandwidth_measurement_bbr_cannot_exit_startup_baseline_fig}
    }
    \subfigure[\name{}]{
    \includegraphics[width=5.7cm, angle=0]{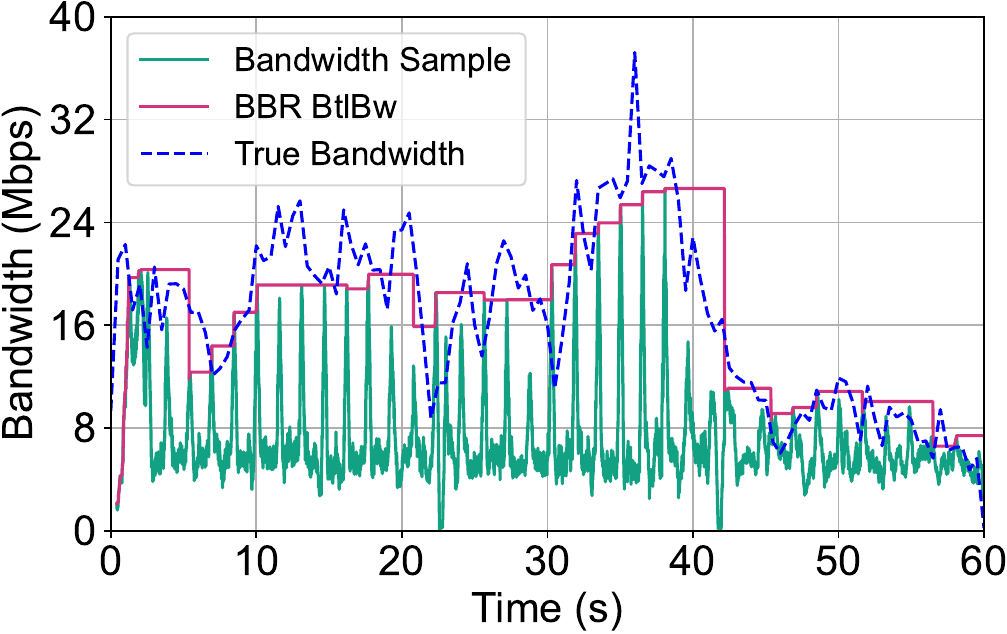}
    \label{bandwidth_measurement_bbr_cannot_exit_startup_bbr_copilot_fig}
    }
        \vspace{-0.2cm}
    \caption{Bandwidth measurement in a network environment where BBR cannot exit the Startup phase.}   \label{bandwidth_measurement_bbr_cannot_exit_startup_fig}
        \vspace{-0.4cm}
\end{figure}

To evaluate the optimization effect of \name{} on addressing issue in the Startup phase of BBR, we need to select a network environment where BBR hardly exits the Startup phase and the buffer size is small. In this case, we evaluate the performance of the \name{} in an environment with a streaming bitrate of 5.4Mbps, RTT of 200ms, network buffer of 40KB, and using the Verizon-LTE-driving.up and Verizon-LTE-driving.down traces provided by Mahimahi to set the uplink and downlink bandwidth. 

As shown in Figure \ref{bandwidth_measurement_bbr_cannot_exit_startup_baseline_fig}, in this environment, for BBR without \name{} deployed, there is sufficient data to be sent within the first 3 seconds. BBR updates its BtlBw to a larger value based on accurate bandwidth measurement samples. After 3 seconds, due to the transport layer's frequent entry into the application-limited phase, BBR without \name{} cannot exit the Startup phase due to a lack of accurate bandwidth measurement samples. Since BBR does not reduce the sending rate during the Startup phase, the sender continues to send data at a rate of pacing\_gain times the BtlBw, resulting in a sending rate of 60.5Mbps, even exceeding the maximum available bandwidth of 37.2Mbps within 60 seconds. With only a 40KB buffer size, a large number of packet losses occur, with a retransmission ratio of 27.5\%. When \name{} is deployed, as shown in Figure \ref{bandwidth_measurement_bbr_cannot_exit_startup_bbr_copilot_fig}, BBR smoothly exits the Startup phase, and the sending rate adjusts according to the true bandwidth, closely matching the true bandwidth. Packet losses are significantly reduced, with a retransmission ratio of 12.6\%.

\begin{figure}[tbp]
    \centering
    \subfigure[Exiting Startup ratio]{
    \includegraphics[width=5.7cm, angle=0]{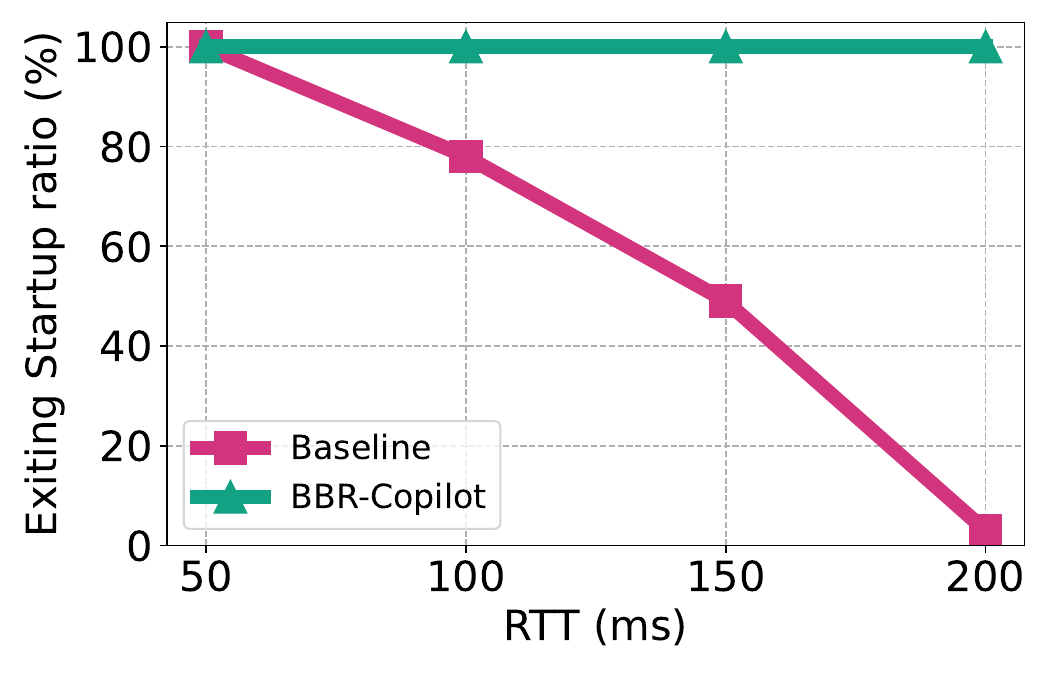}
    \label{exiting_startup_ratio_optimization}
    }
    \subfigure[Retransmission ratio]{
    \includegraphics[width=5.7cm, angle=0]{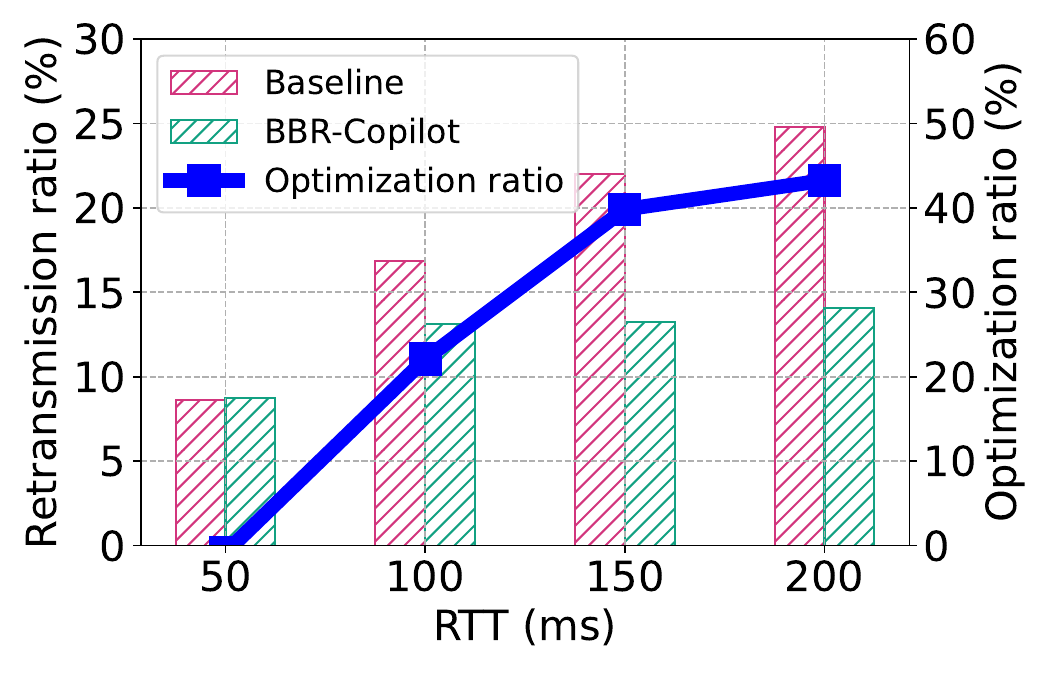}
    \label{retransmission_ratio_optomization_fig}
    }
        \vspace{-0.3cm}
    \caption{The optimization effect of \name{} varies with RTT.}    \label{the_optimization_effect_of_bbr_copilot_fig}
        \vspace{-0.6cm}
\end{figure}

Then, with all other network environment parameters unchanged, we tested the optimization effect of the \name{} on the issue BBR encounters in the Startup phase under environments with RTTs of 50ms, 100ms, 150ms, and 200ms, respectively, to examine how the optimization effect changes with RTT. 

As shown in Figure \ref{exiting_startup_ratio_optimization}, for BBR without \name{} deployed, as the RTT increases from 50ms to 200ms, the proportion of live streams in which BBR exits the Startup phase within 60 seconds decreases from 100\% to 3\%. For BBR with \name{} deployed, regardless of the RTT variation, the proportion of live streams in which BBR exits the Startup phase remains at 100\%. This demonstrates that \name{} effectively helps BBR exit the Startup phase. 

As shown in Figure \ref{retransmission_ratio_optomization_fig}, as the RTT increases from 50ms to 200ms, for BBR without \name{} deployed, the retransmission ratio gradually increases from 8.6\% to 24.8\%. However, for BBR with \name{} deployed, the retransmission ratio increases from 8.7\% to 14.1\%. The optimization ratio relative to BBR without \name{} gradually increases from nearly 0\% to 43.3\%. It is evident that as the RTT increases, the proportion of live streams in which BBR exits the Startup phase decreases, leading to a higher retransmission ratio. \name{} effectively addresses the issue of a high retransmission ratio caused by BBR's failure to exit the Startup phase.

\vspace{-3mm}
\subsection{Optimization of \name{} in The ProbeBW Phase}
\label{evaluation_probebw_sec}
\vspace{-2mm}
\begin{figure}[tbp]
    \centering
    \subfigure[Baseline (RTT = 50ms)]{
    \includegraphics[width=5.7cm, angle=0]{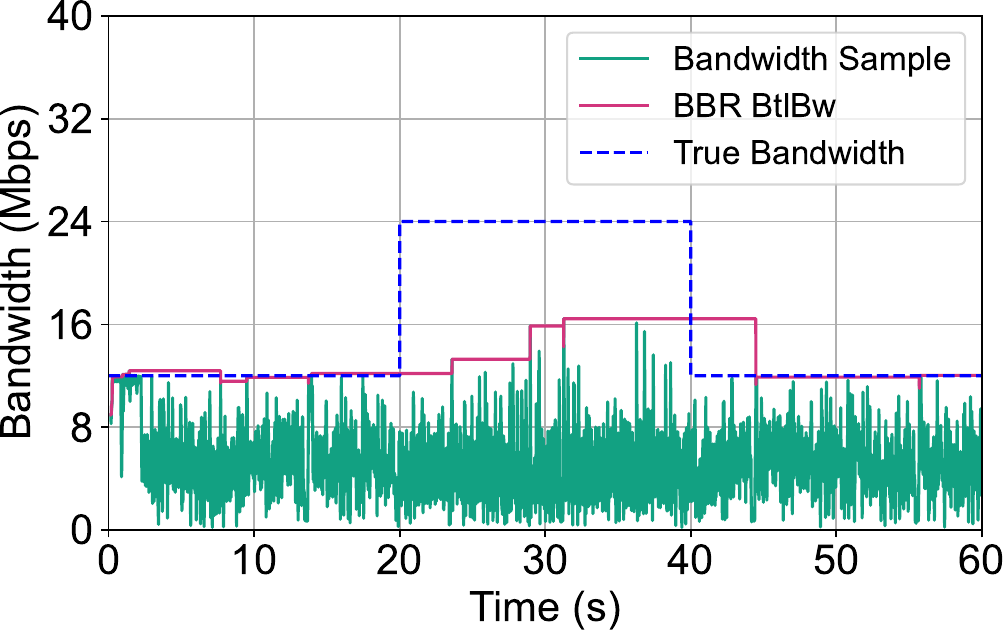}
    \label{bandwidth_measurement_bbr_exit_startup_baseline_delay_50ms_fig}
    }
    \subfigure[\name{} (RTT = 50ms)]{
    \includegraphics[width=5.7cm, angle=0]{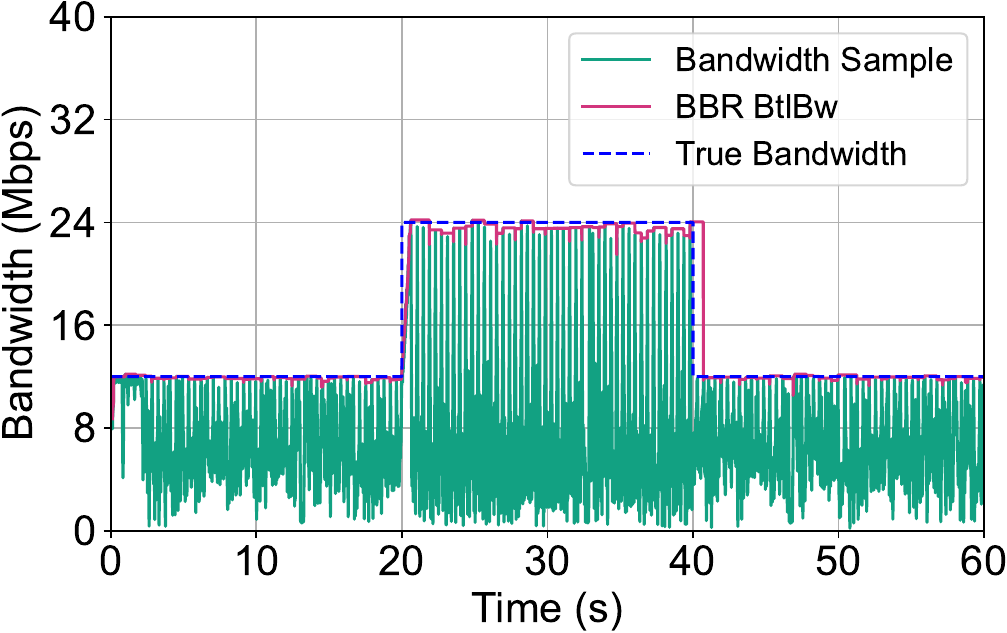}
    \label{bandwidth_measurement_bbr_exit_startup_bbr_copilot_delay_50ms_fig}
    }
        \vspace{-0.3cm}
    \caption{Bandwidth measurement in a low-RTT network environment where BBR can exit the Startup phase.}    \label{bandwidth_measurement_bbr_can_exit_startup_short_rtt_fig}
        \vspace{-0.5cm}
\end{figure}

To evaluate the optimization effect of \name{} on addressing the issue in the ProbeBW phase of BBR, we need to select a network environment where BBR easily exits the Startup phase and the available network bandwidth increases. Here we evaluate the performance of the \name{} in an environment with a streaming bitrate of 5.4Mbps and RTT values of 50ms and 200ms. The uplink bandwidth is set at 12Mbps, while the downlink bandwidth varies as follows: 12Mbps from 0 to 20 seconds, 24Mbps from 20 to 40 seconds, and 12Mbps from 40 to 60 seconds. In this environment, regardless of whether \name{} is deployed or not, BBR can exit the Startup phase and enter the ProbeBW phase after the Drain phase. 

As shown in Figure \ref{bandwidth_measurement_bbr_exit_startup_baseline_delay_50ms_fig}, when the RTT is 50ms, between the 20th and 40th seconds, BBR without \name{} deployed can only obtain a larger bandwidth measurement sample of 16.4Mbps, which is much lower than the true bandwidth of 24.0Mbps, and the RMSE of the true bandwidth and the bandwidth measurement samples is 9.41. As shown in Figure \ref{bandwidth_measurement_bbr_exit_startup_bbr_copilot_delay_50ms_fig}, BBR with \name{} deployed can quickly obtain accurate bandwidth measurement samples that are close to the true bandwidth. The RMSE of the true bandwidth and the bandwidth measurement samples is 1.31 between the 20th and 40th seconds, which is 86.1\% lower than the RMSE of the true bandwidth and the bandwidth measurement samples without \name{} deployed.

\begin{figure}[tbp]
    \centering
    \subfigure[Baseline (RTT = 200ms)]{
    \includegraphics[width=5.7cm, angle=0]{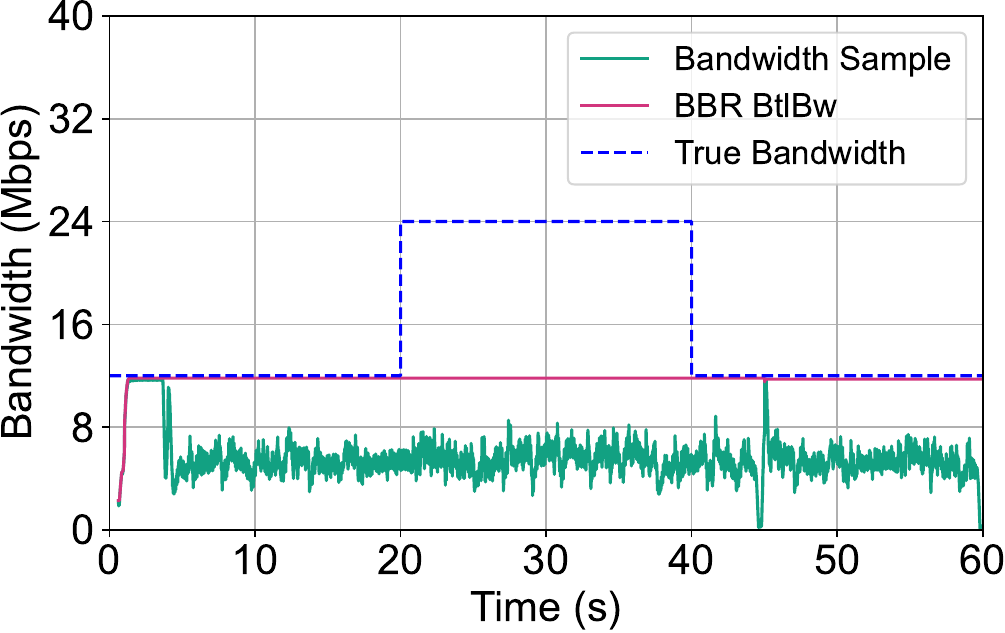}
    \label{bandwidth_measurement_bbr_exit_startup_baseline_delay_200ms_fig}
    }
    \subfigure[\name{} (RTT = 200ms)]{
    \includegraphics[width=5.7cm, angle=0]{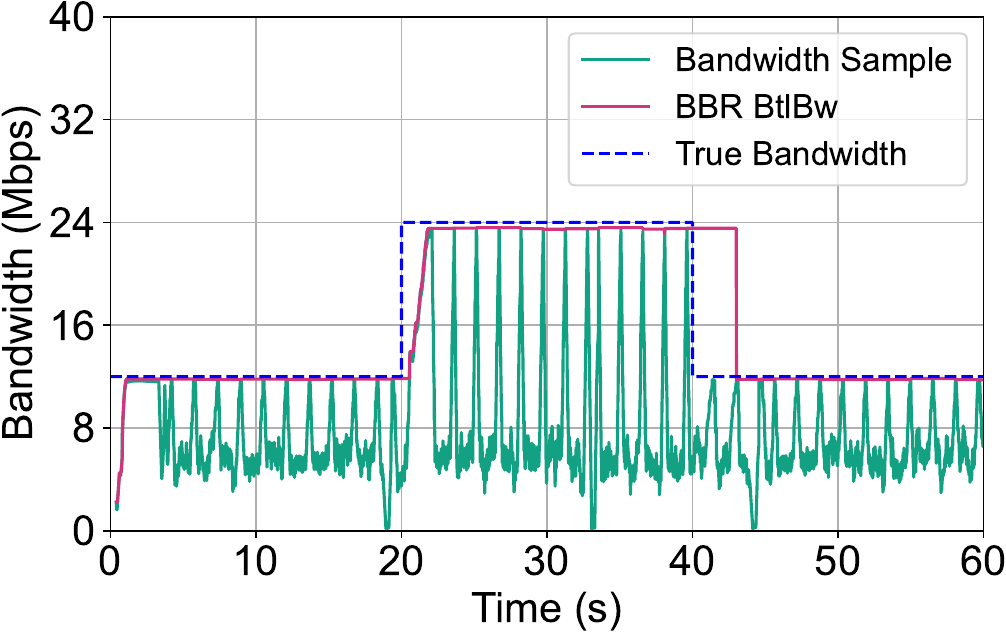}
    \label{bandwidth_measurement_bbr_exit_startup_bbr_copilot_delay_200ms_fig}
    }
        \vspace{-0.2cm}
    \caption{Bandwidth measurement in a high-RTT network environment where BBR can exit the Startup phase.}    \label{bandwidth_measurement_bbr_can_exit_startup_long_rtt_fig}
        \vspace{-0.6cm}
\end{figure}

As shown in Figure \ref{bandwidth_measurement_bbr_exit_startup_baseline_delay_200ms_fig}, when the RTT is 200ms, BBR without \name{} deployed can not even obtain larger bandwidth measurement samples and continues using the bandwidth measurement sample of 11.8Mbps obtained between the 0th and 20th seconds, which results in the RMSE of the true bandwidth and the bandwidth measurement samples being 12.18 between the 20th and 40th seconds. However, as shown in Figure \ref{bandwidth_measurement_bbr_exit_startup_bbr_copilot_delay_200ms_fig}, BBR with \name{} deployed performs better. It can obtain an accurate bandwidth measurement sample exceeding 23.0Mbps. The RMSE of the true bandwidth and the bandwidth measurement samples is 2.72. This is 77.7\% lower than the RMSE of the true bandwidth and the bandwidth measurement samples without \name{} deployed. Therefore, BBR’s sending rate can be more precisely aligned with the available bandwidth.
It is evident that \name{} can effectively improve the accuracy of bandwidth estimation of BBR during the ProbeBW phase, which helps BBR to adjust the sending rate accurately to match the available bandwidth.
\vspace{-3mm}

\vspace{-3mm}

\section{Discussion \& Future Work}
\label{related_work_sec}
\vspace{-4mm}

\vspace{1ex} \noindent \textbf{Extra transmission costs:} Generating extra data to prevent the transport layer from entering the application-limited phase will introduce extra transmission costs. The larger the gap between the true bandwidth and the live-streaming bitrate, the more extra data is needed, and the higher the cost. Therefore, accurately and timely generating accurate bandwidth measurement samples requires the introduction of significant extra transmission costs. The current solution of BBR-Copilot is to generate accurate bandwidth measurement samples only at certain key moments. In future work, we plan to reduce extra transmission costs further, that is, to balance the bandwidth required for live streaming and the extra transmission costs, and choose an appropriate bandwidth as the upper limit for our bandwidth probing, rather than precisely detecting the maximum available bandwidth of the entire link.


\vspace{1ex} \noindent \textbf{Content of extra data:} The extra data currently generated by BBR-Copilot consists of some meaningless filler data.
However, we believe that if we can make full use of this extra data, it can also optimize the transmission of normal data. We plan to utilize copies of in-flight packets as extra data. When we send these extra data to prevent the transport layer from entering an application-limited phase, even if the in-flight data packets are lost, the receiving end can recover the content of these lost packets when it receives these extra data, reducing the time for packet loss recovery.




\vspace{1ex} \noindent \textbf{Higher versions of BBR:} In this paper, we conducted evaluation tests based on version 1 of BBR and did not involve higher versions of BBR. Compared to BBRv1, which only chooses to exit the Startup phase based on estimated BtlBw increase, BBRv2~\cite{cardwell2019bbrv2} and BBRv3~\cite{cardwell2023bbrv3} also choose to exit the Startup phase based on packet loss rate. This may solve the problem of BBRv1 not being able to exit the Startup phase easily in live-streaming scenarios. However, it still cannot solve the problem that BBR cannot adjust the sending rate in time as the actual bandwidth increases in the ProbeBW phase. We plan to continue testing the effects of BBR-Copilot on BBRv2 and BBRv3 in future work.
\vspace{-3mm}

\section{Related Work}
\label{related_work_sec}
\vspace{-3mm}
\vspace{1ex} \noindent \textbf{Congestion control algorithm.}  Currently, widely-deployed congestion control algorithms (CCAs) are mostly rule-based \cite{winstein2013sprout,zaki2015verus,ha2008cubic,CardWell2016bbr}. Designed by domain experts, they adjust packet sending rate and congestion window according to network feedback (e.g., loss, throughput, delay). These algorithms are optimized for specific network environments, yet no single rule-based CCA works universally. This has led to the rise of learning-based CCAs, which use modern machine-learning tools like supervised learning (e.g., Indigo \cite{yan2018pantheon}, Muses \cite{zhong2022muses}) or reinforcement learning (e.g., Aurora \cite{jay2019deep}, DeepCC \cite{abbasloo2020wanna}). They perform well in diverse network scenarios. However, their black-box design causes performance degradation in new environments \cite{zhang2020machine,jiang2021machine}, and high computation overhead restricts large-scale deployment. Recently, the poly-algorithmic congestion control paradigm has been proposed \cite{emara2020eagle,yang2023gemini,zhou2021antelope}, aiming to combine multiple CCAs to leverage strengths and mitigate limitations. Combining rule-based and learning-based CCAs has balanced stability and generality, but the built-in learning-based CCAs still result in high deployment overhead. In summary, compared to other CCAs, BBR performs excellently in common network scenarios, with low overhead and stable performance, making it widely used.

\vspace{1ex} \noindent \textbf{Optimization of BBR.} Since the introduction of BBR, many efforts have been made to optimize the issues inherent in BBR. BBR-S~\cite{chiariotti2021bbrs}, when measuring bandwidth, uses an Adaptive Tobit Kalman Filter (ATKF) to replace the maximum filter that may lead to overestimation of bandwidth, thereby reducing queuing delay. oBBR~\cite{Pengqiang2023obbr} avoids substantial packet loss in shallow buffer links by detecting the size of the bottleneck buffer and adjusting the upper limit of data in transmission. It can also accurately and promptly detect bandwidth drops and adjust its sending rate, significantly reducing the retransmission ratio. BBR-E~\cite{Kim2019bbre} enhances the fairness of flows that share the same bottleneck link but have different RTTs by reducing cwnd when a larger RTT is detected. Unlike previous works, BBR-Copilot does not directly modify BBR. Instead, it adds a mechanism to help BBR more accurately estimate bandwidth in live-streaming scenarios.

\vspace{1ex} \noindent \textbf{On-off traffic pattern.} Many previous works~\cite{pan2023amphis,Wierman2003modeltcpvegas,Kupka2012preformanceofonofftraffic,de2013elastic,zhao2017onofftraffic,yan2023too} have demonstrated the on-off traffic pattern is not conducive to transmission control. The work most closely related to \name{} in recent years is Amphis~\cite{pan2023amphis}. Amphis is a congestion control algorithm framework that enhances the accuracy of bandwidth estimation and the speed of data transmission at the message granularity level under the on-off traffic pattern by dividing the sent data into messages and using FEC coding~\cite{fec2024} to generate extra data to assist in bandwidth estimation. The main differences between BBR-Copilot and Amphis are as follows: (i) We have conducted a detailed analysis of the problems with BBR in the on-off traffic pattern and their causes. (ii) We have provided conditions for generating extra data and the implementation method coupled with BBR in detail. 

\vspace{-3mm}

\section{Conclusion}
\label{conclusion_sec}
\vspace{-2mm}

\name{} assists BBR in obtaining accurate bandwidth measurement samples, enabling it to quickly exit the Startup phase and adjust the sending rate accurately and promptly in the ProbeBW phase. Evaluation results have demonstrated that \name{} significantly enhances the performance of BBR in live-streaming scenarios. In the future, we plan to further reduce the transmission cost associated with \name{} and improve the timeliness of generating accurate bandwidth measurement samples. Subsequently, we will deploy it on our CDN servers, serving millions of users worldwide.


\vspace{-2mm}
\bibliographystyle{splncs04}
\bibliography{reference}
\end{document}